\documentclass[sigconf]{acmart}
\usepackage{xspace,url,xcolor}
\usepackage{subcaption}
\usepackage{listings}
\usepackage{array}
\usepackage{multirow}
\usepackage{tabularx}
\usepackage{listings}
\usepackage[export]{adjustbox}
\usepackage{pifont}
\renewcommand\footnotetextcopyrightpermission[1]{} 
\newcommand{\pgraph}[1]{\smallskip\textbf{#1.}}

\newcommand{\query}{Bew\xspace}
\newcommand{\sys}{BewQA\xspace}

\newcommand{\id}[1]{{\sf #1}}

\newcommand{\myurl}[1]{{\url{#1}}}

\newcommand{\x}{{\sf x}}

\newcommand{\cmark}{\ding{51}}
\newcommand{\eat}[1]{}
\setcopyright{none}

\makeatletter
\def\@maketitle{\newpage
\null
\setbox\@acmtitlebox\vbox{%
\baselineskip 20pt
\vskip 2em                   
  \begin{center}
   {\ttlfnt \@title\par}       
   \vskip 1.5em                
  \end{center}}
\dimen0=\ht\@acmtitlebox
\advance\dimen0 by -7pc\relax 
\unvbox\@acmtitlebox
\ifdim\dimen0<0.0pt\relax\vskip-\dimen0\fi}
\makeatother

\begin{document}
%

\title{Bew: Towards Answering Business-Entity-Related Web Questions	}

\author{%
  Qingqing Cao*,  Oriana Riva
 $^{\dagger}$,  Aruna Balasubramanian*, Niranjan Balasubramanian* }
\affiliation{
 	\institution{*Stony Brook University \quad $^{\dagger}$Microsoft Research \vspace{0.5cm}} 
 }

\begin{abstract}
We present \textit{\sys}, a system specifically designed to answer a class of questions that we call \textit{\query} questions. \query questions are related to businesses/services such as restaurants, hotels, and movie theaters; for example, ``Until what time is happy hour?''. These questions are challenging to answer because the answers are found in open-domain Web, are present in short sentences without surrounding context, and are dynamic since the webpage information can be updated frequently. Under these conditions, existing QA systems perform poorly. We present a practical approach, called \textit{\sys}, that can answer \query queries by mining a template of the business-related webpages and using the template to guide the search. We show how we can extract the template automatically by leveraging aggregator websites that aggregate information about business entities in a domain (e.g., restaurants). We answer a given question by identifying the section from the extracted template that is most likely to contain the answer. By doing so we can extract the answers even when the answer span does not have sufficient context. Importantly, \sys\ does not require any training. We crowdsource a new dataset of 1066 \query questions and ground-truth answers in the restaurant domain. Compared to state-of-the-art QA models, \sys has a 27 percent point improvement in F1 score. Compared to a commercial search engine, \sys answered correctly 29\% more \query questions. 
\end{abstract}

\maketitle

\section{Introduction}
\label{sec:intro}

\begin{figure}[t]
	\centering
  \begin{subfigure}{0.18\textwidth}	
	\centering
     \includegraphics[width=1.01\textwidth]{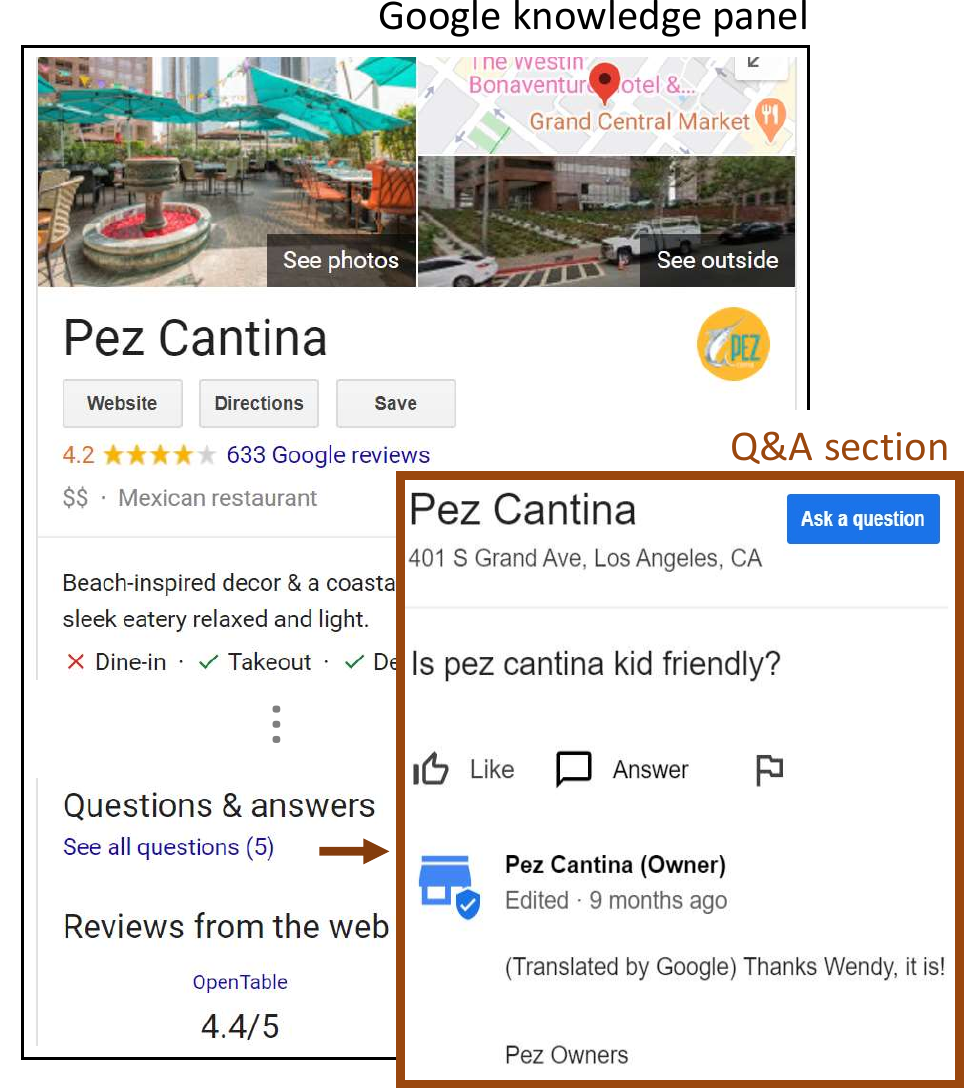}
  \end{subfigure}	
	\hspace{0.05cm}
  \begin{subfigure}{0.28\textwidth}
		\centering
		\scalebox{0.65}{
		\begin{tabular}{lp{1pt}}
		\toprule		
		Question  & G \\
		\midrule					
		How much is the dim sum? & \x \\		
		Is wifi available? Helpful for work lunches  & \x \\
		Do you guys do birthdays? & \x \\ 		
		Are you open tomorrow? & \cmark \\		
		Does anyone know if they cater for external events? & \x \\		
		Do you deliver? & \cmark \\		
		What's dress code for here & \x \\
		Can children eat at this restaurant? & \x \\		
		Do you have access for wheelchairs? & \cmark \\
		Do you offer valet parking? & \x \\
		\bottomrule
		\end{tabular}	
		}
  \end{subfigure}
\caption{The table (on the right) reports 10 restaurant-related questions users asked in Google knowledge panels~\cite{knowledgepanel} to restaurant providers. A knowledge panel with its Q\&A section is shown on the left. We submitted these 10 questions to Google search (G), by appending the associated restaurant names; Google search could answer 3 out 10.}
\label{fig:queries}
\end{figure}





\begin{figure}[t]  
  \includegraphics[width=0.6\linewidth]{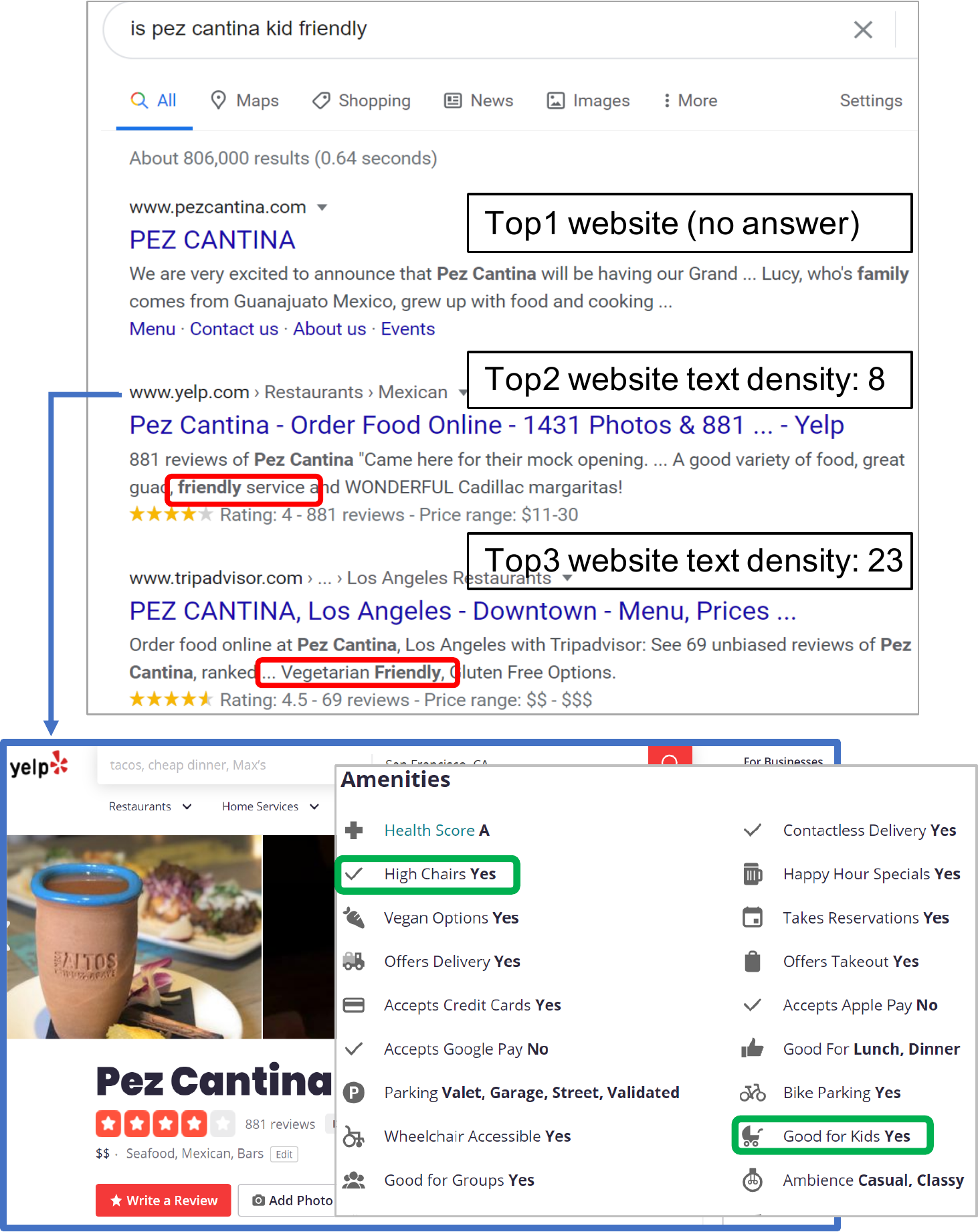}
	\caption{Top 3 search results for the question asked in the knowledge panel of Fig.~\ref{fig:queries}. Google returns no direct answer and also captions are irrelevant. However, the answer exists in 2 of the top 3 search results, but the text density of the answer sections is an order-of-magnitude lower than SQuAD.}
	\label{fig:googleyelp}
\end{figure}

While research on question answering (QA) has made tremendous progress in recent years~\cite{squad}, we identified a large class of questions related to business entities such as restaurants, hotels, movie theaters, etc. which cannot be answered accurately by existing QA systems. Fig.~\ref{fig:queries} lists a sample of questions users ask to restaurant providers in Google; most of them could not be answered automatically by Google search.\footnote{Users asked these questions in Google knowledge panels~\cite{knowledgepanel} for certain restaurant businesses. We submitted the same questions along with the associated restaurant names to Google search. We evaluated whether a direct answer was returned or any relevant highlighted texts appeared in a top search result's caption~\cite{bing-caption}.} The only-answered questions were those that referred to popular attributes of a business entity, such as a restaurant's opening hours, some of which could be retrieved directly from their knowledge graphs. Search engines failed on many questions whose answers were available in the open, unstructured Web. For example, Google search did not return a direct answer to the question ``Is Pez Cantina kid friendly''; but, the webpage linked in the second top search result (Fig.~\ref{fig:googleyelp}) contained the answer.


In this paper, we focus on this new class of web questions related to business entities, which we call \textit{Bew} questions for short. \query questions pose new challenges stemming from three properties of their data sources: \textit{(1) Unstructured Web}: answers to \query questions are in open-domain Web; \textit{(2) Low text-density}: answers are usually present in short sentences or even single words with little (or misleading) surrounding content, as in the example of Fig.~\ref{fig:googleyelp} where the correct answer ``Good for kids'' appears next to ``Bike Parking'' and ``Ambience''. This makes it difficult for text-based QA systems that rely on matching the well-formed textual context for scoring answers~\cite{caiming16,seo17,dcn18}; and \textit{(3) Dynamic}: these documents contain time-sensitive information, hence they are regularly updated; cached indexes get stale quickly, re-indexing on-demand is time consuming, and maintaining knowledge bases may incur high maintenance costs. 


We present a practical approach, called \sys, that can answer \query queries by mining the structure used to semantically organize short-text information in business-related webpages. 
These business webpages follow a \textit{template} and present information in sparse but \textit{informative sections} with distinct titles and consistent layouts that make it easy for users to navigate to the key information quickly.  
Knowledge of this template can help both in scoring relevant business webpages and in identifying correct answers from the selected webpages despite their low text-density content.

To do so, a first challenge \sys must address is how to extract templates for business-related webpages without any explicit supervision or website-specific effort. To this end we leverage \textit{aggregator websites} that aggregate information about business entities in a domain (e.g., restaurants, hotels, tourist attractions, etc.) from various websites, backend APIs, and databases. These aggregator websites typically use consistent layout and styling to present data about various business entities, each in a separate \textit{entity webpage}. For example, Fig.~\ref{fig:googleyelp} shows the webpage of the Pez Cantina restaurant in the aggregator website \myurl{yelp.com}. We leverage this consistent styling to locate section titles and use redundancy to identify important sections using a simple low-effort procedure.

A second challenge \sys must address is how to effectively use templates to (i) identify aggregator websites and sections that are likely to contain correct answers, and (ii) draw correct answers from them. \sys must deal with short texts surrounded by unrelated content, as in the example in Fig.~\ref{fig:googleyelp}. To improve the robustness of this matching, we exploit the redundancy in section information across entity webpages in aggregator websites. Instead of directly asking the entity-specific question against the target business entity webpages, we remove the entity from the question and ask the same question against a pool of webpages of other businesses. This allows us to locate sections and aggregators that are most likely to answer this ``type'' of question.  

While \sys does not require any dataset for training, we do need an evaluation dataset. Since there are no existing datasets for evaluating \query-style queries, we set up a crowdsourcing procedure to create one. We collected 1066 question-answer pairs along with information on sections where these answers appear in aggregator websites. We compared the performance of \sys to that of BERT QA~\cite{bert}, a strong baseline for end-to-end QA, and found \sys has a 27 percentage point improvement in F1 score. \sys was also able to answer 29\% more \query questions than Google Search. 

Overall, this paper makes the following contributions. (i) It introduces a new class of business-related queries that are currently underserved by existing systems. (ii) It proposes a practical solution to answer \query queries effectively based on the notion of aggregator templates. (iii) It proposes a new dataset to study \query queries which will be released to the community. (iv) It evaluates \sys extensively by comparing it to BERT QA models and Google Search.

\section{\query questions}
\label{sec:motivation}
\label{sec:dew_query}


We identify \textbf{\query questions} as a new category of questions which users \textit{already} ask on the Web. These questions are targeted towards a specific service or business, such as a restaurant or hotel. We collected a set of 200 real user questions covering 100 different business entities from the ``Question \& Answer'' sections in Google knowledge panels~\cite{knowledgepanel}. These questions are similar to the 10 shown in Fig.~\ref{fig:queries}. 
We filtered the questions to remove ambiguous, personal questions or ones for which the answer was unavailable on the Web. The remaining 146 in this set are \query questions. These are questions about service/business entities and their answers can be found mainly in the \textit{business-specific website} (e.g., {\em pezcantina.com} for the entity {\em Pez Cantina}) or in \textit{aggregator websites} (e.g., Yelp, OpenTable, etc.). When we run these questions through Google, it returns direct answers for only 25 of them (17\%). In cases where there is no direct answer, if we inspect the first snippet it is possible to locate words related to the answer in an additional 46 cases (31\%). Even so more than half of these \query questions are unanswered.

\pgraph{Why existing QA systems cannot answer \query queries}  \query questions present new challenges for end-to-end question answering. QA systems are often categorized into knowledge-based and IR-based approaches. Knowledge-based systems rely on databases and large-scale knowledge graphs to deduce answers~\cite{berant13,Unger12}. These approaches cannot be used for \query queries as their answers are generally not present in knowledge bases. Further, many of these questions are dynamic in nature and their values can change. A manual inspection of the 146 questions above shows that 96 of them are dynamic, such as ``Who's playing tonight?''  

IR-based systems use a document retrieval module to identify relevant documents and an answer extractor module to draw the answer from them~\cite{chen-etal-2017-reading}. Text-based QA models~\cite{caiming16,seo17,dcn18} rely on the textual context to extract answers -- they rely on correct answers being embedded in surrounding context which matches the information sought in the question. However, this is not the case for \query queries, as the following experiment demonstrates.


We entered each of the 10 questions shown in Fig.~\ref{fig:queries} along with their associated restaurant names in Google search. Google answered only 3 of them. However, when inspecting the webpages linked in the top 3 search results we observed the following. \textit{(i)} \textit{Aggregator websites are a prevalent source of correct answers to \query questions}. For 9 out of the 10 questions, at least one of the top 3 webpages contained a correct answer; in 7 cases, the answers were present in aggregator websites, while for 2 questions the answers were found in the business-specific website. \textit{(ii)} \textit{Answers to \query questions appear in low text-density webpages.} We divided each webpage into informative sections (we discuss this methodology in \S\ref{sec:method}) and computed the text density~\cite{text-density} for each section. The average text density of the question-relevant sections across the aggregator webpages was 21.3, and for the business-specific webpages was 16.2. As an alternative, we sampled 10 factoid questions from the SQuAD\cite{squad} dataset and similarly looked at the top 3 webpages returned by Google. The text density was much higher, at 183.2.

\section{System design}
\label{sec:method}

This section describes the challenges that \sys needs to address to answer \query queries and how its design deals with them.

\subsection{Intuition and challenges}

A key challenge with \query questions is that their answers appear in low text-density webpages, which makes it hard to build adequate context to identify them. On the other hand, we observe that, partly because of their low text-density property, to be human-readable these webpages tend to follow a \textit{template} consisting of various \textit{informative sections}, each with a distinct title and with a predefined location within the template. \sys automatically extracts templates for these webpages and leverages them to (i) retrieve webpages which are most relevant to a question, and (ii) influence answer scoring with section scoring. \sys is designed to be a \textit{widely-applicable} and \textit{practical} system. It uses three key ideas.

{\bf Extracting templates automatically}. Existing work that leverages structural information of a webpage assumes that the content is organized in tables~\cite{pasupat2015compositional,Sun16} or that is annotated with structured data (e.g., \id{Schema.org}~\cite{schema2qa}). We make neither of these assumptions, as they would limit the applicability of \sys. For example, we found in the \query dataset (\S\ref{sec:task}) that the \id{Schema.org} structured data only covers 23.4\% of the section properties that provide correct answers. Instead, we leverage \textit{aggregator websites} which, as shown in \S\ref{sec:motivation}, are often top-ranked sources for \query questions. Aggregator websites aggregate information about entities in a domain (restaurants, hotels, tourist attractions, etc.) from various sources, such as websites, APIs, and databases. They typically use consistent layout and styling to present data about various entities. For example, for the restaurant domain, an aggregator website such as \myurl{opentable.com} contains various {\em entity} webpages for different restaurants, all organized using a similar template. We show in \S\ref{sec:aggregator} how these aggregator websites can be used to automatically infer the semantic structure of a webpage describing a certain entity.

{\bf Retrieving entity webpages effectively}. Answers to \query queries can change frequently (e.g., holiday hours, deals, ratings, availability, etc.) and a user may ask these questions for many a large number of entities including newly-listed or less-frequent ones (e.g., a new restaurant or an unusual attraction). In general, we cannot assume an extensive and up-to-date collection of all webpages related to a certain entity to be available. 
Moreover, scoring individual entity webpages against a question can produce noisy results due to their low text-density. For all these reasons, we score a pool of cached entity webpages (generally unrelated to the entity of interest) from various aggregator websites and use this assessment to decide which aggregator websites are most relevant to a question. Webpages related to the entity of interest are fetched \textit{only} from those few selected aggregators (more details in \S\ref{sec:matcher}). 

{\bf Scoring answers with robustness}. Even after identifying an aggregator webpage that contains the correct answer, we can still fail in extracting it because it is short-text and surrounded by unrelated content. For instance, imagine scoring an answer to a question about ``wheelchair access'' from a section titled ``Amenities'' as in Fig.~\ref{fig:googleyelp}. The words ``wheelchair accessible'' are surrounded by unrelated words which can cause answer extraction to focus on the wrong parts of the section and draw an incorrect answer, even when the answer extractor is based on an advanced neural comprehension model~\cite{jia-liang-2017-adversarial} (see experiment with BERT-QA~\cite{bert} in \S\ref{sec:abl}). To cope with this issue, we score the sections of \textit{other} entity webpages with respect to the \textit{same} question and score the entity-related answers higher if they come from highly-ranked sections.

\begin{figure}[t]
\centering
\includegraphics[width=\columnwidth]{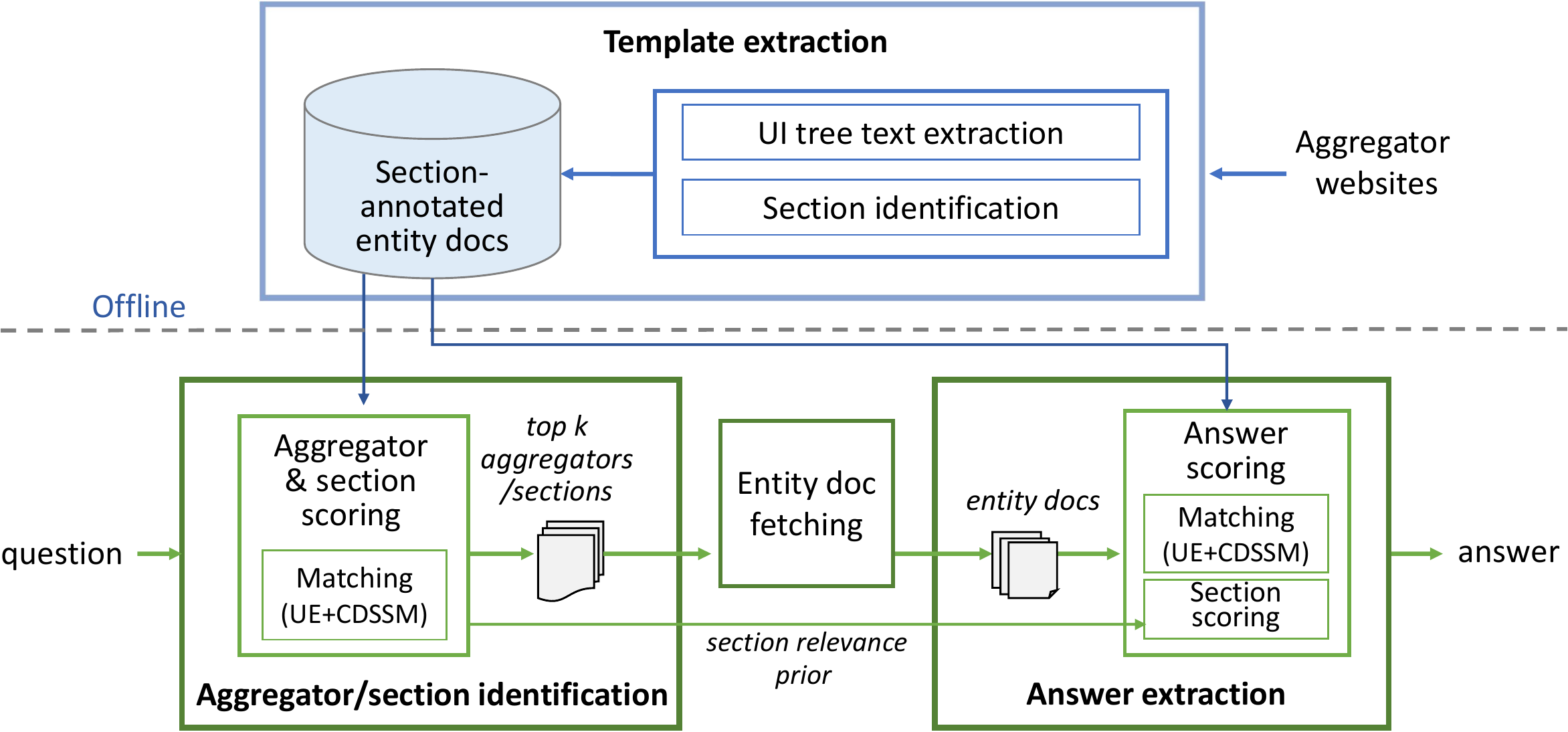}
\caption{\sys architecture.}
\label{fig:arch}
\end{figure}

Aligned with these three ideas, Fig.~\ref{fig:arch} shows the architecture of \sys. Offline, it extracts a template for each aggregator website. During runtime, it uses the templates to rank the aggregators and their sections without considering the target entity, and then looks for these sections in the entity webpages and scores answers.

\subsection{Template extraction}
\label{sec:aggregator}

\begin{figure}[t]
	\centering
	\includegraphics[width=\linewidth]{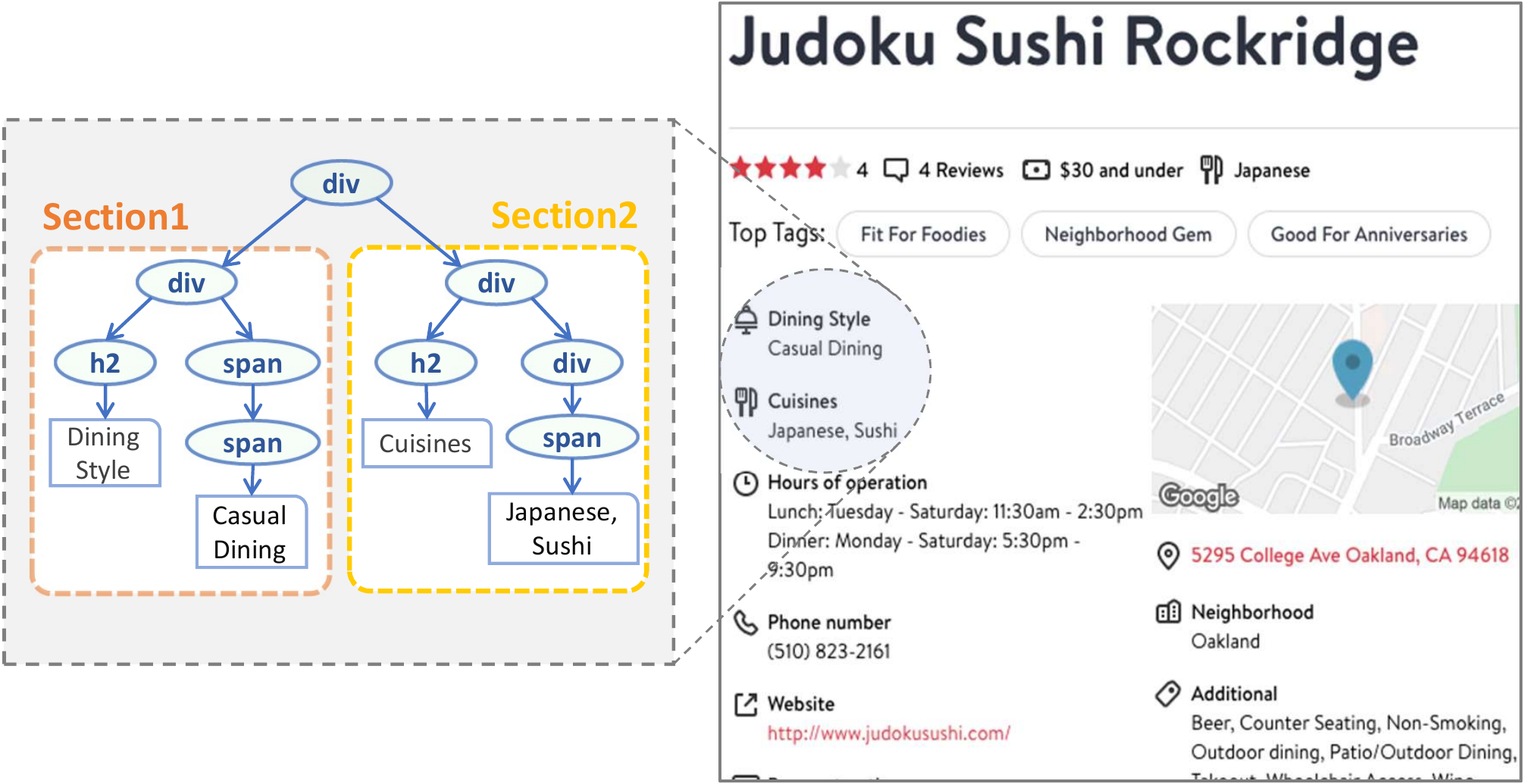}
	\caption{A restaurant entity website in the aggregator website \myurl{opentable.com} (right) and a portion of its UI tree showing the layout and dependency of the ``Dining Style'' and ``Cuisines'' sections (left).}
	\label{fig:dom}
\end{figure}

The goal of this offline process is to associate with each aggregator website a {\em template} which captures the main informative sections that are common across the different entity webpages listed by the aggregator. A {\em template} is a list of phrases that denote {\em sections} of content that contain semantically-related texts with consistent styling. Fig.~\ref{fig:dom} shows the entity webpage for the entity {\em Jodoku Sushi Rockridge} within the aggregator website {\em opentable.com}. {\em Hours of operation} and {\em Phone number} are examples of sections. The main challenge is knowing which parts of the webpage provide important information about an entity, without requiring manual inspection. 

A benefit of using aggregator websites is that they present information in a consistent fashion.
Most entity webpages under an aggregator website organize information under sections with same or similar titles
that are styled consistently using large bold fonts to indicate salience. We exploit the consistent styling to locate section titles and uses word frequency to identify important sections, without requiring any explicit supervision or manual effort.

To this end, we create templates by analyzing a small set of entity webpages (e.g., 100 webpages) from each aggregator website. First, we remove non-relevant content including ads, banners, and copyright using simple heuristics. 
Then, we render these webpages in a browser to get the full UI tree (the DOM tree) to get all the text, including those from dynamic elements. We traverse the UI tree to extract text and styling information for all textual nodes using the CSS styling attributes 
and build a frequency map of the extracted texts. 
We then mark the most-frequently appearing texts as \textit{immutable texts} and further mark those immutables with title-like styling as \textit{title immutables}.

The last stage in this process is to identify a section's boundaries based on the identified section titles. Starting from the title text node, we traverses the UI tree in a bottom-up fashion and find the maximum subtree that contains at least another text node and that does not contain any other section title. For example, in Fig.~\ref{fig:dom}, we start from the ``Dining Style'' node classified as title and go up one level to find the first \id{div} node. If we go up one more level, the next \id{div} node has a subtree containing another title node, hence we stop here. The first-encountered \id{div} node defines the first section titled ``Dining Style''. Similarly, we identify the second section ``Cuisines''.

\subsection{Aggregator and section scoring}
\label{sec:matcher}

Given a question $q$ related to an entity $e$, we need to determine which aggregator websites and which sections within them are likely to provide correct answers. Simply scoring all entity webpages for $e$ against $q$ may not be possible nor advantageous for two reasons. First, the entity webpages for $e$ may have not been fetched yet (because $e$ is not popular) or they may contain stale content. Second, aggregator websites are low text-density, hence directly scoring their templates and short-text contents against $q$ may produce low-accuracy results. For example, a section titled ``Additional'' that lists miscellaneous items (Fig.~\ref{fig:dom}) is ambiguous. 

To deal with these constraints we exploit the redundancy in the aggregator websites. Imagine we are given the question ``Does $e$ have live music today?". Based on a sample of webpages related to other entities it can be inferred that an aggregator website $a$ typically list information relevant to the question in the sections with the title ``Entertainment''. This suggests that we are more likely to find the answer in the ``Entertainment'' section in the entity webpage related to $e$ from $a$. We operationalize this idea as below to compute an \textit{aggregator-level score} and \textit{section-level score} that we then use to boost the scores of the answers.

To score each aggregator website, we use a small sample of $N$ entity webpages (e.g., N=20) from each aggregator website. We remove any entity mentions from the question by means of simple grammatical rules and a repository of entity names. For example, the question ``When does Altura open?'' is transformed into ``When does \id{it} open?''. Then, we semantically score the entity-agnostic question with respect to all sections across all webpages. For each section, we consider the title, the body's texts and structured data (if any) such as \id{Schema.org} annotations. Structured data annotations are appended to the section's texts and can help boost the scoring accuracy if the section's body consists of short texts. 


To compute the semantic similarity between a question and a section we use encodings from two models: (i) Universal Encoder (UE)~\cite{cer2018universal}, a Transformer-based model which has been shown to produce reliable encodings for phrase-level texts appropriate for low-density texts, and (ii) CDSSM~\cite{cdssm}, a convolutional-network based model that is trained specifically for information retrieval objectives.\footnote{We use pre-trained models and do not apply any training procedures.} Empirically, we find that CDSSM-based scoring tends to favor literal matching (e.g., direct word-level and character-level matches) while UE scoring allows for matches in high-level semantics. Because of their complementary characteristics, combining the two encoding models provides higher accuracy compared to using them individually (see \S\ref{sec:abl}). For each section, we tokenize\footnote{We use the spaCy tokenizer from \myurl{https://github.com/explosion/spaCy}.} the text from the section into phrases and obtain a simple cosine similarity between the question and section encoded using CDSSM and UE (separately). The final \textit{section relevance score} is a weighted average of these two similarity scores for all phrases in the section.

For a question $q$ the section relevance score for a section $s$ is computed as follows:
\begin{align}
match(q, t_i) &= \frac{sim(UE(q), UE(t_i)) + sim(CDSSM(q), CDSSM(t_i))}{2} \nonumber \\
secScore(q, s) &= \sum^{n}_{i=1} match(q, t_i) * \frac{match(q, t_i)}{\sum^{n}_{i=1} match(q, t_i)} \label{eqn:sec-score}
\end{align}

where, $match(q, t_i)$ denotes the scoring function between $q$ and the $i$-th phrase of $s$ ($n$ phrases in total) and is computed as an average of the cosine similarity ($sim$) of the UE and CDSSM models. 

We then use these section relevance scores to compute a relevance score for the aggregator website as a whole -- i.e., how likely it is for the entity webpage from this aggregator website to contain the answer. To this end, we first compute the \textit{page relevance score} as the weighted average of the top-$k$ section scores in each entity webpage, and the \textit{aggregator relevance score} as the average of the top-$k$ page relevance scores.

Given N entity webpages $\{p_1, ..., p_N\}$ for aggregator $a$, we compute the relevance score for each webpage and the aggregator as:
\begin{align}
pageScore(q, p_i) &= \sum_{j = 1}^{k} secScore(q, s_j) * \frac{secScore(q, s_j)}{\sum^{k}_{j=1} secScore(q, s_j)}\\
aggScore(q, a) &= \frac{1}{N} \sum_{i = 1}^{N} pageScore(q, p_i)
\end{align}

\subsection{Answer scoring}
\label{sec:ans-ext}

To find answers, we only retrieve the $e$-related webpages in the top $M$ (e.g., M= 5) aggregator websites based on the aggregator relevance scores ($aggScore$) defined in the previous subsection. For each such webpage, we first identify the top-$k$ sections that are likely to contain a correct answer. Then, we extract and score answer candidates from all of these sections.

To identify the top-$k$ sections we score each section's text and structured data (if any) as well as a prior that indicates how likely it is for the answer to be found in a section of that type. Given an $e$-related webpage $d_i$ from an aggregator $a$, for each section $s_{ij}$ in $d_i$, we first compute the section score $secScore(q, s_{ij})$ as shown in Eq.~\ref{eqn:sec-score}, using the same procedure as discussed in the previous subsection. The only difference is that the sections $s_{ij}$ now come from an $e$-related webpage, one that is about the entity in question. To each of these section scores we also add a {\em section relevance prior}, $secPrior(q, s_{ij})$, based on the section scores we computed in the previous phase using randomly-sampled entity webpages from the same aggregator $a$. The intuition in doing this is that an assessment of the section relevance based on a pool of documents larger than one (i.e., the entity-related webpage only) can further help with the low text-density challenge by boosting the most-relevant sections (see an experiment on this in \S\ref{sec:abl} for more details). Let $W_{ij}$ denote the sections in the random N webpages from the aggregator $a$ that have the same section title as $s_{ij}$. We compute this prior as the average of the scores of these $W_{ij}$ sections. The final score for the section $s_{ij}$ is the sum of the direct section relevance score and the section relevance prior for the section as shown below:
\begin{align}
prior(q, s_{ij}) &= \frac{1}{|W_{ij}|} \sum_{w \in W_{ij}} secScore(q, w) \\
secFinalScore(q, s_{ij}) &= secScore(q, s_{ij}) + secPrior(q, s_{ij}) \label{eqn:proir-score}
\end{align}

To extract and score answers, we consider candidate answers from all sections in all $e$-related webpages. The section's texts are tokenized to obtain candidate answer phrases. 
Given a section $s_k$, a candidate phrase $c_{kl}$ is scored using the matching function in Eq.~\ref{eqn:sec-score}. To favor answers coming from sections that were previously identified as relevant, we boost the matching-based score with the section relevance score. Formally, the answer relevance score for each candidate answer phrase is computed as follows:
\begin{equation}
ansScore(q, c_{kl}) = match(q, c_{kl}) + secFinalScore(q, s_{k})
\end{equation}
We return a ranked list of candidate answers according to this score.

\section{Evaluation Dataset}
\label{sec:task}
\label{sec:anl}

To evaluate \sys we require real user queries and ground-truth answers. One option is to collect question-answer pairs from knowledge panels of search engines, discussion forums or Q\&A sections of aggregator websites. These sources are, however,  unreliable as their contributors may provide wrong or out-of-date answers, or refer to content that is not available online (thus hindering automatic evaluation). It is also challenging to automatically map a user answer to specific texts or passages in a webpage.  

Instead, we create a new dataset of crowdsourced questions and answer pairs from crowd workers. We focus the data collection on the restaurant domain because many restaurant aggregator websites exist and users are familiar with them. Overall, our goal is to collect questions that users are likely to ask from different webpages, along with the answer span on the webpage. In all, we collect a dataset of 1066 question-answer pairs from 124 webpages for 26 entities, and we use this for our evaluation. 

\pgraph{Crowdsourcing goals and challenges} We select 14 popular aggregator websites for restaurants, including \href{https://www.opentable.com/}{OpenTable}, \href{https://www.yelp.com/}{Yelp}, \href{https://www.tripadvisor.com/}{TripAdvisor}, and \href{https://www.zagat.com/}{Zagat}. We randomly select 26 restaurant entities that have a corresponding webpage in at least three aggregator sites, 
and collect 124 entity webpages across all the aggregator sites. 

Our goal is to crowdsource questions pertaining to each entity and ground-truth answer spans. This goal entails two challenges. First, we need to crowdsource high-quality questions that relate to a given business entity. Second, we need to find ground-truth answers in \textit{all} the corresponding pages about the business entity across the aggregator websites. This is important for automatically evaluating the answers returned by \sys, since the system can find the answer in any one of the aggregator webpages.  

To this end, we conduct two rounds of Human Intelligence Tasks (HITs) on Mechanical Turk~\cite{turk}. In the first round, we collect user questions about the entities. In the second round, we obtain the answer spans for the questions in all available entity webpages.



\begin{figure*}[t]
	\centering
  \begin{subfigure}{0.35\textwidth}	
	\centering
     \includegraphics[width=\linewidth]{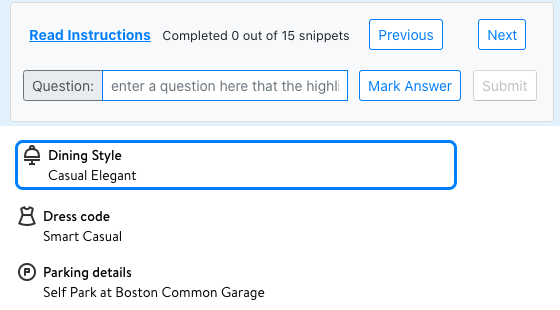}
			\caption{First round}
			\label{fig:first-round}
  \end{subfigure}	
	\hspace{3ex}
  \begin{subfigure}{0.37\textwidth}
		\centering
			\includegraphics[width=\linewidth]{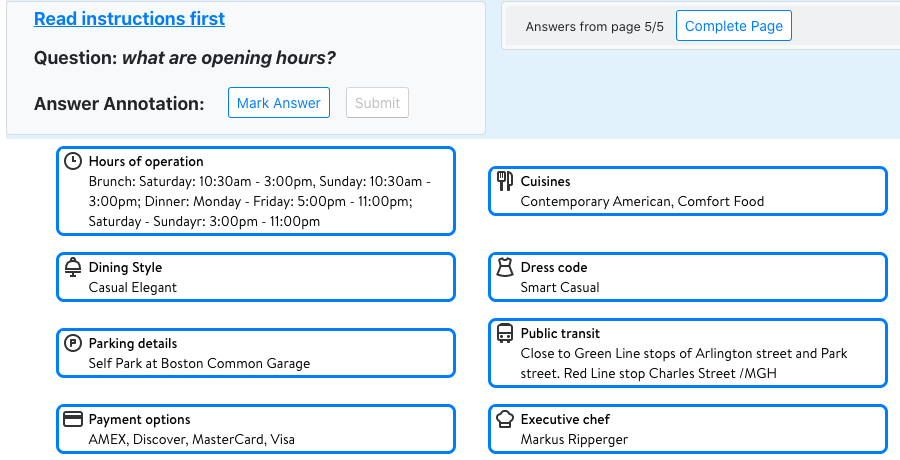}
			\caption{Second round}
			\label{fig:second-round}
  \end{subfigure}
\caption{HIT interfaces. In the first round, workers enter a question for each highlighted section snippet and label the answer in the snippet. In the second round, given an entity-related question, they label the answer for all highlighted snippets for all available (aggregator) webpages for that entity. }
\label{fig:hit-interfaces}
\end{figure*}

\pgraph{Crowdsourcing methodology} In the first round task, we collect questions for all the entities. For each entity, we randomly select one entity webpage from any of the aggregator websites. Using a webpage snapshot tool~\cite{singlefile}, we save a rendered copy of the page as a single HTML file and divide it manually into section snippets. Fig.~\ref{fig:first-round} shows the HIT interface for the task. 
The task is to propose a question for each highlighted snippet and annotate the answer span. For example, for the highlighted snippet in the figure, the worker may type the question ``Is there a dress code?'', select the text ``Casual Elegant'' and click the ``Mark Answer'' button. Workers can navigate through the snippets, pick one, pose a question, and annotate the answer span. After doing all snippets, the ``Submit'' button is enabled. We asked the workers to read detailed instructions on how to perform the task including a short video demonstrating it. Workers were paid based on the number of annotated snippets, and no more than 3 workers could annotate the same webpage. 

In this task, we had 82 workers and collected 1276 question-answer pairs. Additionally, for quality control, we recruited three graduate students at a local university to inspect the collected questions and then either discard ``bad'' questions or change them into ``good'' questions. The criteria for bad or good questions was explained as follows. A bad question is vague in the information asked (e.g., ``About the restaurant?'') and/or irrelevant to the highlighted snippet (e.g., ``It is a modern restaurant?'' asked for the {\em ratings} snippet). By doing this we obtained 1066 good-quality questions.  


In the second round, given a question, the workers select a snippet matching it and label the answer span (see Fig.~\ref{fig:second-round}). The answer must be either in one of the highlighted snippet or non-existent (empty label). For each question, the task is repeated for all entity webpages across all aggregator sites. The same question was assigned to a maximum of 3 workers. If workers marked different answer spans for the same question, the answer with the highest consensus (if any) was kept. In all, we had 68 workers annotating 124 webpages. 
For quality control, we checked whether the answer span from the first round's workers and the second round's workers matched. Overall, we discarded 13\% of non-matching answers. 

\pgraph{Dataset} We collected 1066 questions for 26 entities over 124 webpages in 14 aggregator websites. Table~\ref{tab:stats} summarizes the dataset.

\begin{table}[t]
    \centering
    \small
    \caption{Statistics of \query dataset.}
    \label{tab:stats}
    \begin{tabular}{ll}
        \toprule
         \# questions & 1066 \\
         \# entities & 26 \\
         \# aggregator websites & 14 \\
				 \# entity webpages & 124 \\
         \# annotators per query & 2.3 \\
         Avg (min,max) \# sections per webpage & 8.2 (3,19) \\
         Avg (min,max) \# answers per question & 4.7 (1,7) \\
         Mean (min,max) question words & 4.9 (2,12) \\
				 Mean (min,max) section words & 13.4 (2,248) \\
         Mean (min,max) answer words & 3.2 (2,35) \\
         \bottomrule
    \end{tabular}
\end{table}

\section{Evaluation}
\label{sec:evaluation}

The goal of this section is to answer the following questions: {\bf (i)} How does \sys perform on the collected dataset and how does it compare to other baselines? {\bf (ii)} How does \sys compare to deployed QA systems such as Google search?, and {\bf (iii)} Which components of \sys have a major impact on its performance?


\subsection{Methodology}

\paragraph{Dataset}
We evaluate the performance of \sys and baseline systems on the \query dataset (\S\ref{sec:task}). Since \sys does not require training, we use the entire dataset (1066 questions) for evaluation.

\paragraph{Baseline systems}

\newcommand{\docqa}{Doc-BERT-QA\xspace}
\newcommand{\irqa}{IR-BERT-QA\xspace}
\newcommand{\google}{Google Search\xspace}

We compare the performance of \sys against \docqa, \irqa, and \google.

{\bf \docqa} leverages BERT~\cite{bert} to extract answers from entity webpages. BERT has achieved impressive results for many NLP tasks; we simply fine-tuned it on the SQuAD~\cite{squad} dataset for QA purposes. We tried fine-tuning BERT on the \query dataset, but this resulted in a 39\% drop in performance (possibly due to overfitting). 
To feed web documents to BERT, we obtain the text blocks from the webpages by accessing the corresponding UI tree and concatenate them into paragraphs.

{\bf \irqa} uses an information retrieval (IR) approach to identify relevant text blocks in webpages and then extracts answers from those text blocks. 
As before, we obtain the text blocks from the webpage UI tree and build the webpage index. We then use BM25 term weighting search\footnote{We used the Python wrapper (PyLucene) of the Lucene search engine.} 
to retrieve the best-match text nodes and invoke the BERT QA model to extract answers from those texts.

As both \docqa and \irqa do not depend on aggregator-derived templates, for them we use both aggregator websites {\em and} the business-specific websites as input for answer extraction (for \sys we use only aggregator websites).

For {\bf \google} we submit the queries to \google and inspect whether a direct answer is returned. In the absence of a highly-confident answer, \google sometimes highlights words related to the questions in the captions of the top search results. We do not classify them as ``answers'' because a user still needs to parse the full caption to extract them and because multiple, possibly unrelated, words can be highlighted in the same caption.

\paragraph{Answer evaluation}

For \sys, \docqa, and \irqa, we evaluate the extracted answer strings by automatically comparing them against the ground-truth answers in our dataset. To compare \sys\ and \google's answers, since we do not know which web documents \google uses 
to answer the submitted queries, we adopt a manual judgement process. Specifically, we asked 3 subjects to manually inspect \google's direct answers and decide whether they are appropriate answers for the submitted questions; otherwise, the subject reports that there is no answer. We kept judgements that had at least two agreements. Because this comparison requires manual judgement, we only perform this evaluation with a subset of 100 \query questions. 

\begin{table}[t!]
	\centering
	\setlength\tabcolsep{3pt}
	\caption{Performance comparison of \sys, \docqa and \irqa over the \query dataset (n=1066 questions).}
	\begin{tabular}{@{}lcccccc@{}}
		\toprule
		Method & F1@1 & F1@2 & F1@3 & EM@1 & EM@2 & EM@3 \\ 
		\midrule
		\docqa & 0.28 & 0.34 & 0.41 & 0.20  & 0.25 & 0.31 \\
		\irqa & 0.36 & 0.50  & 0.53 & 0.17 & 0.24 & 0.30  \\
		\textbf{\sys}   & \textbf{0.63}  & \textbf{0.73}  & \textbf{0.79}  & \textbf{0.36}  & \textbf{0.45}   & \textbf{0.48}  \\ 
		\bottomrule
		\end{tabular}	
	\label{table:qa-results}
\end{table}



\begin{table}[t!]
	\centering
	\setlength\tabcolsep{7pt}
	\small	
	\caption{Comparison of \sys, \google and \docqa for 100 randomly-selected \query questions.}
	\begin{tabular}{@{}lcccc@{}}
		\toprule
		Method  & Correct & Wrong & No Answer & Accuracy \\ \midrule
		\google & 36      & \textbf{7}     & 57        & 0.36     \\
		\docqa  & 50      & 50    & \textbf{0}         & 0.50     \\
		\sys    & \textbf{65}      & 35    & \textbf{0}         & \textbf{0.65}     \\ \bottomrule
		\end{tabular}	
	\label{tab:google}
\end{table}

\begin{table*}[ht]
	\hspace{-0.8cm}
	\begin{minipage}[t]{0.35\linewidth}
		\centering
		\small	
		\caption{Example \query questions where \google gives wrong direct answers and \sys outputs correct answers.}
		\begin{tabular}{l}\toprule
			\textbf{Example question}  \\
			\textbf{\google answer (G)} | \textbf{\sys answer (B)} \\			
			\midrule
			Q=Is table reservation required of Canlis?\\		
			G: Need to know | B: Table reservation required \\
			\midrule
			Q=Type of cuisines in 520 Bar Grill \\
			G: Grilled hanger steak | B: American \\
			\midrule
			Q=Is there parking at Maggiano's Little Italy? \\
			G: map widget (no parking info) | B: parking available \\ 
			\bottomrule
		\end{tabular}
		\label{tb:eg-queries}
	\end{minipage}
	\hspace{0.8cm}
	\begin{minipage}[t]{0.55\linewidth}
		\centering
		\small	
		\setlength\tabcolsep{2pt}
		\caption{Ablation analysis over 160 \query queries. The table reports F1 score, EM and section precision and whether the correct (C) or wrong (W) answer (A) came from a correct or wrong section (S) (e.g., ``CS-WA'' means the answer is wrong but came from a correct section.)}
		\begin{tabular}{@{}lccccccc@{}}
			\toprule
												 & F1@1 & EM@1 & Sec-P@1 & CS-WA & CS-CA & WS-CA & WS-WA \\ \midrule
			\sys    & 0.76 & 0.53 & 0.85 & 0.32 & 0.53 & 0 & 0.15 \\ 
			\midrule
			\quad CDSSM-only match & 0.60 & 0.43 & 0.68 & 0.25 & 0.43 & 0 & 0.33 \\
			\quad UE-only match & 0.70 & 0.44 & 0.79 & 0.35 & 0.44 & 0 & 0.21 \\ 		
			\quad answer scoring w/o secPrior & 0.57 & 0.37 & 0.70 & 0.34 & 0.36 & 0.01 & 0.29 \\
			\quad answer scoring w/ BERT-QA & 0.30 & 0.21 & 0.18 & 0.04 & 0.14 & 0.06 & 0.76 \\
			\bottomrule
			\end{tabular}
		\label{table:ablation}
	\end{minipage}
\end{table*}


\paragraph{Metrics} For the evaluation of all systems except \google, we compute exact match (EM) and F1 scores. Exact match measures the percentage of predictions that match any one of the ground-truth answers exactly. 
F1 score measures the average word-level overlap between the prediction and ground-truth answer. We treat the prediction and ground truth as bags of tokens, and compute their F1 score. We take the maximum F1 score over all of the ground-truth answers for a given question. 
To get a deeper understanding of the system performance, for both metrics we report \textbf{EM@k} and \textbf{F1@k} by considering the top $k$ results, with $k$=1,2,3.

When evaluating \sys it is important to consider also how precisely the system identifies the informative sections relevant to a question. In the \query dataset, in addition to the ground-truth answers we have collected the ground-truth sections. Therefore, for \sys, we also compute precision and recall in identifying sections in entity webpages; \textbf{sec-P@k} is the section precision by considering the top $k$ results.


To compare the performance of \sys\ and \google, our manual judgement process labels each question as correctly or wrongly answered. Hence, we compute the \textbf{accuracy} as the number of correct answers divided by the total questions. 

\subsection{Overall results}

\pgraph{Baseline comparison on \query dataset} We ran \sys, \docqa and \irqa over all the 1066 \query questions. As Table \ref{table:qa-results} reports, \sys outperformed both \docqa and \irqa in EM@1 by 16 points and 19 points, respectively, and also in terms of F1@1 scores (35 points and 27 points higher, respectively). This better performance attributes to the fact that \sys can correctly identify sections containing correct answers. A correct section was ranked top in 71\% of the cases (or 85\% if considering the top-3 ranked sections). By identifying the sections correctly, \sys\ is able to find the answer more accurately even when the text-density is low. 

\pgraph{Comparison with Google Search} We conducted a small-scale experiment to evaluate how current search engines such as \google perform on \query queries. Based on 100 randomly-selected \query questions, \sys's accuracy was 29 points and 15 points higher than that of \google and \docqa, respectively (see Table~\ref{tab:google}). Interestingly, \google yielded a high precision of 84\% (36/43) and a relatively low recall (43\%). This is because \google returns a direct answer only when it has high confidence in its correctness, but high-confidence cases are not common in general. Table~\ref{tb:eg-queries} shows three examples where \google gave a wrong direct answer (while \sys identified a correct answer). 

\pgraph{Section identification} We measured the performance of \sys in correctly identifying sections in an entity webpage, both in the offline and online phases. We used all the entity webpages in the \query dataset. Overall, \sys accurately identified the sections yielding {\bf 81\%} precision and {\bf 88\%} recall for the online phase and {\bf 84\%} precision and {\bf 93\%} recall for offline phase. 	

\subsection{Ablation analysis of \sys}
\label{sec:abl}

\paragraph{\sys techniques} We evaluate the effectiveness of three main techniques used in \sys: {\bf matching function} (Eq.~\ref{eqn:sec-score} in \S\ref{sec:matcher}), {\bf answer scoring} and {\bf answer extraction}. We varied these three techniques as follows. (i) We replaced \sys's matching function with two matching functions that use CDSSM and Universal Encoder (UE) individually (instead of combining them as we do). (ii) When extracting the final answers, we scored answers without boosting the scores using the section relevance prior scores, i.e. $secPrior(q, s_{ij})$ in Eq.~\ref{eqn:proir-score}. 
(iii) We replaced \sys's answer scoring approach with BERT-QA (i.e., BERT-QA extracts answers from the sections \sys identified in the entity webpages).

Table~\ref{table:ablation} reports the results on 160 randomly-selected questions from the \query dataset. Overall, section re-ranking gave the most benefits (removing this component caused 19 points drop in F1@1) showing it is key to boost sections containing the correct answers. Replacing our semantic matching approach caused F1@1 to drop by 6 points (UE) and by 16 points (CDSSM). This is likely because multiple semantic matchers pre-trained on different datasets provide an ensemble that better captures the semantics of short texts.

\begin{table*}[htb!]
	\footnotesize
	\caption{We manually categorized the 149 erroneous (F1@1 is 0) predictions in the 1066 \query questions. 
	}
	\begin{tabular}{p{1.39cm} p{6cm} p{8.7cm} r}
		\toprule
	{\bf Category}    & {\bf Description}   & {\bf Example}  & {\bf \%} \\
	\midrule
	Matching bias & Semantic matching assigns high similarity scores to salient domain words (e.g. food, bar, address, etc.) & Question: \textit{What's the address?} -- Correct answer: \textit{10146 Main St, Bellevue, WA} (score: 0.41) \newline Selected answer: \textit{Email} (score: 0.52 -- ``email'' is semantically close to ``address''). & 43\% \\ \midrule
	False sections & A subtree in the page UI tree is wrongly classified as a section due to website inconsistent DOM styling, hidden content, etc. & The text label appearing next to a drop-down menu in the webpage is classified as section title due to the lack of meaningful DOM attributes and complex subtree structure.  
	& 26\% \\ \midrule
	UI tree \newline misplace & Webpage developers put texts associated with a section outside of the section's UI subtree & An ``Opening hours'' section is located in the UI subtree of an ``Address'' section.  & 23\% \\  \midrule
	False answers & Ground truth answers are either wrong or missing & Annotators marked ``Find a table'' instead of ``Make a reservation'' as the answer to ``Can I make a reservation online?''  &  8\% \\
	\bottomrule
	\end{tabular}
	\label{table:error-analysis}
	\vspace{-0.1in}
	\end{table*}

Interestingly, using the BERT QA model to extract answers made the performance drop (46 points degradation in F1@1). This happened mainly because the BERT QA model could not identify the document sections containing the correct answers -- in 76\% of cases BERT extracted a wrong answer from a wrong section (WS-WA rate). BERT tended to favor answers from relatively-long sections (words $>20$) over possibly correct answers from shorter sections. 



\subsection{\sys error analysis}
Table~\ref{table:error-analysis} presents a taxonomy of the main causes for incorrectly-answered questions by \sys. The analysis is based on a sample of 149 failures over the entire \query dataset. The majority (43\%) of failure cases can be attributed to matching bias, which affects both the section identification and answer extraction phases. Fine-tuning semantic matchers on a domain-specific dataset might help alleviate this problem. The false section identification and UI tree misplace errors are mostly due to flawed or heavily customized website designs. A more robust section identification algorithm with a trained neural network might be more tolerant to these website structure problems. 
Additionally, we notice that ground-truth annotations are not perfect (they caused 8\% of the errors) due to wrong or missing answers. Applying stronger quality control and providing better incentives to crowdworkers might mitigate the erroneous labels.

\section{Related Work}
\label{sec:related}

Web QA systems can be classified into three main categories, depending on the data source used to extract answers: \textit{(i)} structured data, such as knowledge bases or databases, \textit{(ii)} semi-structured data, such as web tables or developer-annotated web schemas, and \textit{(iii)} unstructured data, such as text passages from Wikipedia or news articles.

{\bf QA using structured data.}
These systems parse natural language questions to build a formal semantic representation of the query such as logic forms, graph queries, and SPARQL queries, which are used to query knowledge bases~\cite{Zhang16,Unger12,Yahya12,Zou14}. Transforming the queries often requires sophisticated query understanding techniques and populating such knowledge bases is hard to scale. Structuring the knowledge extracted from aggregator websites is particularly high-effort because aggregators use different terms and formats to represent similar information and constantly update it. In general, by not tying the implementation to a specific schema, we can provide a more flexible a scalable solution.

{\bf QA using semi-structured data.} Web documents are semi-structured in nature; DOM trees annotations, web tables and schema markups such as \id{Schema.org} are examples of web metadata. Schema2QA~\cite{schema2qa} relies on \id{Schema.org} markups to build virtual assistance skills and answer compositional queries. However, studies~\cite{microsoftPrioritizeSearchBoost2017} report that only 17\% of marketers use \id{Schema.org} markup, thus making this approach less applicable. Moreover, ontologies like \id{Schema.org} are not comprehensive enough to accurately annotate all different types of webpages. We found in the \query dataset, the \id{Schema.org} structured data only cover 23.4\% of the section properties containing correct answers. Similarly, table-based QA systems~\cite{pasupat2015compositional,Sun16,vakulenkoS17,webtables} identify the cell(s) in a web table answering a given question. We share with these systems the idea of using entity types or a table schema as valuable clues for answering questions. However, in our case, the schema is not given but instead needs to be inferred. Early work like QuASM~\cite{pinto02} exploits the DOM attributes and style information such as headings to segment a document into smaller blocks (text snippets and tables), and treats these as text documents for QA. QuASM uses various heuristics to chunk webpages, whereas \sys proposes a more robust approach to obtain document templates and extract answers.

{\bf QA using unstructured data.} To handle unstructured data, many QA systems adopt a two-step process. First, the relevant documents and text passages are retrieved using IR techniques. Then, reading comprehension models are used to identify the span of text in the passage or document that best represents an answer. IR-based QA systems have been around for decades~\cite{Jurafsky00,Voorhees00,Ko07,Brill02,Schlaefer06,Leidner04}.  Previous work like POIReviewQA~\cite{mai2018POIReviewQASemantically} extracts point of interest (POI) types from unstructured sentences on the web page and uses a Lucence-based IR method to support open-domain search and QA over geo spatial content. Recent work AmazonQA~\cite{lipton2019AmazonQAReviewBased} studies QA over Amazon product reviews. The reviews section in the web pages are high text-density document. Other recent approaches such as REALM~\cite{realm}, DrQA~\cite{drqa}, ORQA~\cite{orqa} adopt advanced neural machine comprehension models. As discussed earlier (and shown in our evaluation), these systems cannot answer \query queries effectively due to the low text-density of the business webpages that contain answers to \query queries.

{\bf Short text ranking for QA.} Previous work~\cite{annm,short-text-cnn} studies neural matching methods to rank short answer sentences. However, they focus on ranking well-formed and unstructured short texts and their techniques work for open-domain or microblog domains. Instead, \sys solves distinctive challenges in business servicing domains where the answer documents are semi-structured, more dynamic and low text-density. Beyond QA, another thread of research related to \sys is short text retrieval~\cite{fast-short-text}. These systems propose a fast approach to access a small subset of short text candidates, but require building appropriate indices offline. The web documents to answer \query questions are dynamic and need online processing.

{\bf Identifying structure in web documents.} A large body of research has focused on mining webpages with the goal to automatically identify the most information-rich blocks in a webpage (e.g., removing ads, branding banners, footers, etc.) as well as to extract a template structure~\cite{Cai03,Song04,Debnath05,fernandes2007ComputingBlock,kohlschutter2010BoilerplateDetection,vieira2006FastRobust}. This work is particularly important to enhance the performance of search engines in classifying web documents~\cite{vieira2006FastRobust,kim2011TEXTAutomatic}. Most of these techniques learn from collections of documents with the same templates by exploring repetitions in tree patterns, tag sequences, word patterns, and visual features~\cite{Lin02,Yi03,Chen06,kim2011TEXTAutomatic,vieira2006FastRobust}. In some cases, page-level template detection has also been considered by training a classifier to assign ``templateness'' scores to each node in a DOM tree~\cite{chakrabarti2007PagelevelTemplate,Debnath05}. Our approach is unsupervised and rely on a small collection of webpages for each aggregator site. Moreover, while much of this work has been designed for news articles and has focused on the top-level page layout, we focus on the fine-grained semantic structures (section titles and keywords) inside the main body of a webpage. 



\section{Conclusions}
\label{sec:conclusions}

Addressing business-related information needs is important both for the users and the businesses that serve them. 
Yet, these remain underserved by existing QA systems. In this work, we introduced \query questions that are targeted towards businesses. The main difficulty is that information is consistently expressed in low text-density sections in business entity webpages making it difficult for existing QA systems to effectively match surrounding context to score answers. We turn this difficulty to our advantage and show how we can leverage these repeating structures to identify most informative sections for answering a query. We introduced \sys an unsupervised, practical, and scalable QA system that mines the repeating information in aggregator websites to extract templates and informative sections and uses these to locate answers effectively. Evaluations show that this approach outperforms a standard text-based solution that uses a state-of-the-art QA system. We release our \query questions dataset to further research in understanding and improving QA systems for business-related information needs.


%

\bibliographystyle{abbrvnat}
\scriptsize{
  \bibliography{ref}

\begin{thebibliography}{51}
\providecommand{\natexlab}[1]{#1}
\providecommand{\url}[1]{\texttt{#1}}
\expandafter\ifx\csname urlstyle\endcsname\relax
  \providecommand{\doi}[1]{doi: #1}\else
  \providecommand{\doi}{doi: \begingroup \urlstyle{rm}\Url}\fi

\bibitem[{Amazon}(2019)]{turk}
{Amazon}.
\newblock Mechanical turk.
\newblock https://www.mturk.com, 2019.

\bibitem[Berant et~al.(2013)Berant, Chou, Frostig, and Liang]{berant13}
J.~Berant, A.~Chou, R.~Frostig, and P.~Liang.
\newblock Semantic parsing on {F}reebase from question-answer pairs.
\newblock In \emph{Proc. of the 2013 Conference on Empirical Methods in Natural
  Language Processing}, pages 1533--1544. ACL, Oct. 2013.

\bibitem[{Bing blogs}(2010)]{bing-caption}
{Bing blogs}.
\newblock {Anatomy of a Bing Caption}.
\newblock
  \url{https://blogs.bing.com/webmaster/2010/10/25/anatomy-of-a-bing-caption},
  2010.

\bibitem[Brill et~al.(2002)Brill, Dumais, and Banko]{Brill02}
E.~Brill, S.~Dumais, and M.~Banko.
\newblock An analysis of the {{AskMSR}} question-answering system.
\newblock In \emph{Proc. of the {{ACL}}-02 Conference on Empirical Methods in
  Natural Language Processing - Volume 10}, {{EMNLP}} '02, pages 257--264,
  {Stroudsburg, PA, USA}, 2002. {Association for Computational Linguistics}.

\bibitem[Cafarella et~al.(2008)Cafarella, Halevy, Wang, Wu, and
  Zhang]{webtables}
M.~J. Cafarella, A.~Halevy, D.~Z. Wang, E.~Wu, and Y.~Zhang.
\newblock {WebTables: Exploring the Power of Tables on the Web}.
\newblock \emph{Proc. VLDB Endow.}, 1\penalty0 (1):\penalty0 538–549, Aug.
  2008.
\newblock ISSN 2150-8097.

\bibitem[Cai et~al.(2003)Cai, Yu, Wen, and Ma]{Cai03}
D.~Cai, S.~Yu, J.-R. Wen, and W.-Y. Ma.
\newblock Extracting content structure for web pages based on visual
  representation.
\newblock In \emph{Proc. of the 5th Asia-Pacific Web Conference on Web
  Technologies and Applications}, {{APWeb}}'03, pages 406--417, {Xian, China},
  2003. {Springer-Verlag}.
\newblock ISBN 3-540-02354-2.

\bibitem[Cer et~al.(2018)Cer, Yang, Kong, Hua, Limtiaco, John, Constant,
  {Guajardo-Cespedes}, Yuan, Tar, et~al.]{cer2018universal}
D.~Cer, Y.~Yang, S.-y. Kong, N.~Hua, N.~Limtiaco, R.~S. John, N.~Constant,
  M.~{Guajardo-Cespedes}, S.~Yuan, C.~Tar, et~al.
\newblock Universal sentence encoder.
\newblock \emph{arXiv preprint arXiv:1803.11175}, 2018.

\bibitem[Chakrabarti et~al.(2007)Chakrabarti, Kumar, and
  Punera]{chakrabarti2007PagelevelTemplate}
D.~Chakrabarti, R.~Kumar, and K.~Punera.
\newblock Page-level template detection via isotonic smoothing.
\newblock In \emph{Proceedings of the 16th international conference on {World}
  {Wide} {Web}}, {WWW} '07, pages 61--70, New York, NY, USA, May 2007.
  Association for Computing Machinery.
\newblock ISBN 978-1-59593-654-7.
\newblock \doi{10.1145/1242572.1242582}.
\newblock URL \url{https://doi.org/10.1145/1242572.1242582}.

\bibitem[Chen et~al.(2017{\natexlab{a}})Chen, Fisch, Weston, and
  Bordes]{chen-etal-2017-reading}
D.~Chen, A.~Fisch, J.~Weston, and A.~Bordes.
\newblock Reading {W}ikipedia to answer open-domain questions.
\newblock In \emph{Proc. of the 55th Annual Meeting of the Association for
  Computational Linguistics (Volume 1: Long Papers)}, pages 1870--1879.
  Association for Computational Linguistics, July 2017{\natexlab{a}}.

\bibitem[Chen et~al.(2017{\natexlab{b}})Chen, Fisch, Weston, and Bordes]{drqa}
D.~Chen, A.~Fisch, J.~Weston, and A.~Bordes.
\newblock Reading wikipedia to answer open-domain questions.
\newblock \emph{arXiv preprint arXiv:1704.00051}, 2017{\natexlab{b}}.

\bibitem[Chen et~al.(2006)Chen, Ye, and Li]{Chen06}
L.~Chen, S.~Ye, and X.~Li.
\newblock Template detection for large scale search engines.
\newblock In \emph{Proc. of the 2006 {{ACM}} Symposium on Applied Computing},
  {{SAC}} '06, pages 1094--1098, {Dijon, France}, 2006. {ACM}.
\newblock ISBN 1-59593-108-2.

\bibitem[Debnath et~al.(2005)Debnath, Mitra, Pal, and Giles]{Debnath05}
S.~Debnath, P.~Mitra, N.~Pal, and C.~L. Giles.
\newblock Automatic identification of informative sections of web pages.
\newblock \emph{IEEE Trans. on Knowl. and Data Eng.}, 17\penalty0 (9):\penalty0
  1233--1246, Sept. 2005.
\newblock ISSN 1041-4347.

\bibitem[Devlin et~al.(2019)Devlin, Chang, Lee, and Toutanova]{bert}
J.~Devlin, M.-W. Chang, K.~Lee, and K.~Toutanova.
\newblock {{BERT}}: {{Pre}}-training of deep bidirectional transformers for
  language understanding.
\newblock In \emph{Proc. of the 2019 Conference of the North American Chapter
  of the Association for Computational Linguistics: {{Human}} Language
  Technologies, Volume 1 (Long and Short Papers)}, pages 4171--4186, 2019.

\bibitem[Fernandes et~al.(2007)Fernandes, de~Moura, Ribeiro-Neto, da~Silva, and
  Gonçalves]{fernandes2007ComputingBlock}
D.~Fernandes, E.~S. de~Moura, B.~Ribeiro-Neto, A.~S. da~Silva, and M.~A.
  Gonçalves.
\newblock Computing block importance for searching on web sites.
\newblock In \emph{Proceedings of the sixteenth {ACM} conference on
  {Conference} on information and knowledge management}, {CIKM} '07, pages
  165--174, New York, NY, USA, Nov. 2007. Association for Computing Machinery.
\newblock ISBN 978-1-59593-803-9.
\newblock \doi{10.1145/1321440.1321466}.
\newblock URL \url{https://doi.org/10.1145/1321440.1321466}.

\bibitem[Google(2020)]{knowledgepanel}
Google.
\newblock {About Knowledge Panels -- Knowledge Panel Help}.
\newblock \url{https://support.google.com/knowledgepanel/answer/9163198}, 2020.

\bibitem[Gu et~al.(2016)Gu, Yang, Zhou, Qu, Wei, and Shi]{fast-short-text}
Y.~Gu, Z.~Yang, J.~Zhou, W.~Qu, J.~Wei, and X.~Shi.
\newblock A fast approach for semantic similar short texts retrieval.
\newblock In \emph{Proc. of the 54th Annual Meeting of the Association for
  Computational Linguistics (Volume 2: Short Papers)}, pages 89--94, Berlin,
  Germany, Aug. 2016. Association for Computational Linguistics.

\bibitem[Guu et~al.(2020)Guu, Lee, Tung, Pasupat, and Chang]{realm}
K.~Guu, K.~Lee, Z.~Tung, P.~Pasupat, and M.-W. Chang.
\newblock Realm: {{Retrieval}}-augmented language model pre-training.
\newblock \emph{arXiv preprint arXiv:2002.08909}, 2020.

\bibitem[Jia and Liang(2017)]{jia-liang-2017-adversarial}
R.~Jia and P.~Liang.
\newblock Adversarial examples for evaluating reading comprehension systems.
\newblock In \emph{{Proc. of the 2017 Conference on Empirical Methods in
  Natural Language Processing}}, pages 2021--2031, Copenhagen, Denmark, Sept.
  2017. Association for Computational Linguistics.

\bibitem[Jurafsky and Martin(2000)]{Jurafsky00}
D.~Jurafsky and J.~H. Martin.
\newblock \emph{Speech and Language Processing: {{An}} Introduction to Natural
  Language Processing, Computational Linguistics, and Speech Recognition
  (Question Answering Chapter)}.
\newblock {Prentice Hall PTR}, {Upper Saddle River, NJ, USA}, 1st edition,
  2000.
\newblock ISBN 0-13-095069-6.

\bibitem[Kim and Shim(2011)]{kim2011TEXTAutomatic}
C.~Kim and K.~Shim.
\newblock {TEXT}: {Automatic} {Template} {Extraction} from {Heterogeneous}
  {Web} {Pages}.
\newblock \emph{IEEE Transactions on Knowledge and Data Engineering},
  23\penalty0 (4):\penalty0 612--626, Apr. 2011.
\newblock ISSN 1558-2191.
\newblock \doi{10.1109/TKDE.2010.140}.
\newblock Conference Name: IEEE Transactions on Knowledge and Data Engineering.

\bibitem[Ko et~al.(2007)Ko, Nyberg, and Si]{Ko07}
J.~Ko, E.~Nyberg, and L.~Si.
\newblock A probabilistic graphical model for joint answer ranking in question
  answering.
\newblock In \emph{Proc. of the 30th Annual International {{ACM SIGIR}}
  Conference on Research and Development in Information Retrieval}, {{SIGIR}}
  '07, pages 343--350. {ACM}, 2007.
\newblock ISBN 978-1-59593-597-7.

\bibitem[Kohlschütter et~al.(2010)Kohlschütter, Fankhauser, and
  Nejdl]{kohlschutter2010BoilerplateDetection}
C.~Kohlschütter, P.~Fankhauser, and W.~Nejdl.
\newblock Boilerplate detection using shallow text features.
\newblock In \emph{Proceedings of the third {ACM} international conference on
  {Web} search and data mining}, {WSDM} '10, pages 441--450, New York, NY, USA,
  Feb. 2010. Association for Computing Machinery.
\newblock ISBN 978-1-60558-889-6.
\newblock \doi{10.1145/1718487.1718542}.
\newblock URL \url{https://doi.org/10.1145/1718487.1718542}.

\bibitem[Lee et~al.(2019)Lee, Chang, and Toutanova]{orqa}
K.~Lee, M.-W. Chang, and K.~Toutanova.
\newblock Latent retrieval for weakly supervised open domain question
  answering.
\newblock In \emph{Proc. of the 57th Annual Meeting of the Association for
  Computational Linguistics}, pages 6086--6096, 2019.

\bibitem[Leidner(2004)]{Leidner04}
J.~L. Leidner.
\newblock Open-domain question answering from large text collection, m. {{Pa{\c
  S}Ca}}.
\newblock \emph{J. of Logic, Lang. and Inf.}, 13\penalty0 (3):\penalty0
  373--376, June 2004.
\newblock ISSN 0925-8531.

\bibitem[Lin and Ho(2002)]{Lin02}
S.-H. Lin and J.-M. Ho.
\newblock Discovering informative content blocks from web documents.
\newblock In \emph{Proc. of the Eighth {{ACM SIGKDD}} International Conference
  on Knowledge Discovery and Data Mining}, {{KDD}} '02, pages 588--593,
  {Edmonton, Alberta, Canada}, 2002. {ACM}.
\newblock ISBN 1-58113-567-X.

\bibitem[Lipton et~al.(2019)Lipton, Gupta, Kulkarni, Chanda, and
  Rayasam]{lipton2019AmazonQAReviewBased}
Z.~C. Lipton, M.~Gupta, N.~Kulkarni, R.~Chanda, and A.~Rayasam.
\newblock {AmazonQA}: {A} {Review}-{Based} {Question} {Answering} {Task}.
\newblock pages 4996--5002, 2019.
\newblock URL \url{https://www.ijcai.org/Proceedings/2019/0694}.

\bibitem[Lormeau(2019)]{singlefile}
G.~Lormeau.
\newblock {{SingleFile}}.
\newblock https://github.com/gildas-lormeau/SingleFile, 2019.

\bibitem[Mai et~al.(2018)Mai, Janowicz, He, Liu, and
  Lao]{mai2018POIReviewQASemantically}
G.~Mai, K.~Janowicz, C.~He, S.~Liu, and N.~Lao.
\newblock {POIReviewQA}: {A} {Semantically} {Enriched} {POI} {Retrieval} and
  {Question} {Answering} {Dataset}.
\newblock In \emph{Proceedings of the 12th {Workshop} on {Geographic}
  {Information} {Retrieval}}, {GIR}'18, pages 1--2, New York, NY, USA, Nov.
  2018. Association for Computing Machinery.
\newblock ISBN 978-1-4503-6034-0.
\newblock \doi{10.1145/3281354.3281359}.
\newblock URL \url{https://doi.org/10.1145/3281354.3281359}.

\bibitem[Microsoft and Catalyst(2017)]{microsoftPrioritizeSearchBoost2017}
Microsoft and Catalyst.
\newblock Prioritize search to boost marketing {{ROI}}.
\newblock Technical report, 2017.

\bibitem[Pasupat and Liang(2015)]{pasupat2015compositional}
P.~Pasupat and P.~Liang.
\newblock Compositional semantic parsing on semi-structured tables.
\newblock \emph{arXiv preprint arXiv:1508.00305}, 2015.

\bibitem[Pinto et~al.(2002)Pinto, Branstein, Coleman, Croft, King, Li, and
  Wei]{pinto02}
D.~Pinto, M.~Branstein, R.~Coleman, W.~B. Croft, M.~King, W.~Li, and X.~Wei.
\newblock {{QuASM}}: {{A}} system for question answering using semi-structured
  data.
\newblock In \emph{Proc. of the {{2Nd ACM}}/{{IEEE}}-{{CS}} Joint Conference on
  Digital Libraries}, {{JCDL}} '02, pages 46--55, {Portland, Oregon, USA},
  2002. {ACM}.
\newblock ISBN 1-58113-513-0.

\bibitem[Rajpurkar et~al.(2016)Rajpurkar, Zhang, Lopyrev, and Liang]{squad}
P.~Rajpurkar, J.~Zhang, K.~Lopyrev, and P.~Liang.
\newblock {{SQuAD}}: 100,000+ questions for machine comprehension of text.
\newblock In \emph{Proc. of the 2016 Conference on Empirical Methods in Natural
  Language Processing}, pages 2383--2392, 2016.

\bibitem[Schlaefer et~al.(2006)Schlaefer, Gieselmann, Schaaf, and
  Waibel]{Schlaefer06}
N.~Schlaefer, P.~Gieselmann, T.~Schaaf, and A.~Waibel.
\newblock A pattern learning approach to question answering within the ephyra
  framework.
\newblock In \emph{Proc. of the 9th International Conference on Text, Speech
  and Dialogue}, {{TSD}}'06, pages 687--694. {Springer-Verlag}, 2006.
\newblock ISBN 3-540-39090-1 978-3-540-39090-9.

\bibitem[Seo et~al.(2017)Seo, Kembhavi, Farhadi, and Hajishirzi]{seo17}
M.~J. Seo, A.~Kembhavi, A.~Farhadi, and H.~Hajishirzi.
\newblock Bidirectional attention flow for machine comprehension.
\newblock In \emph{Proc. of the 5th International Conference on Learning
  Representations}, {ICLR '17}, 2017.

\bibitem[Severyn and Moschitti(2015{\natexlab{a}})]{annm}
A.~Severyn and A.~Moschitti.
\newblock Learning to rank short text pairs with convolutional deep neural
  networks.
\newblock In \emph{Proc. of the 38th International ACM SIGIR Conference on
  Research and Development in Information Retrieval}, SIGIR ’15, page
  373–382, New York, NY, USA, 2015{\natexlab{a}}. Association for Computing
  Machinery.
\newblock ISBN 9781450336215.

\bibitem[Severyn and Moschitti(2015{\natexlab{b}})]{short-text-cnn}
A.~Severyn and A.~Moschitti.
\newblock Learning to rank short text pairs with convolutional deep neural
  networks.
\newblock In \emph{Proc. of the 38th International ACM SIGIR Conference on
  Research and Development in Information Retrieval}, SIGIR ’15, page
  373–382, New York, NY, USA, 2015{\natexlab{b}}. Association for Computing
  Machinery.
\newblock ISBN 9781450336215.

\bibitem[Shen et~al.(2014)Shen, He, Gao, Deng, and Mesnil]{cdssm}
Y.~Shen, X.~He, J.~Gao, L.~Deng, and G.~Mesnil.
\newblock A latent semantic model with convolutional-pooling structure for
  information retrieval.
\newblock In \emph{Proc. of the 23rd {{ACM}} International Conference on
  Conference on Information and Knowledge Management}, pages 101--110. {ACM},
  2014.

\bibitem[Song et~al.(2004)Song, Liu, Wen, and Ma]{Song04}
R.~Song, H.~Liu, J.-R. Wen, and W.-Y. Ma.
\newblock Learning block importance models for web pages.
\newblock In \emph{Proc. of the 13th International Conference on World Wide
  Web}, {{WWW}} '04, pages 203--211, {New York, NY, USA}, 2004. {ACM}.
\newblock ISBN 1-58113-844-X.

\bibitem[Sun et~al.(2011)Sun, Song, and Liao]{text-density}
F.~Sun, D.~Song, and L.~Liao.
\newblock {{DOM}} based content extraction via text density.
\newblock In \emph{Proc. of the 34th International {{ACM SIGIR}} Conference on
  Research and Development in Information Retrieval}, {{SIGIR}} '11, pages
  245--254. {ACM}, 2011.
\newblock ISBN 978-1-4503-0757-4.

\bibitem[Sun et~al.(2016)Sun, Ma, He, Yih, Su, and Yan]{Sun16}
H.~Sun, H.~Ma, X.~He, W.-t. Yih, Y.~Su, and X.~Yan.
\newblock Table cell search for question answering.
\newblock In \emph{Proc. of the 25th International Conference on World Wide
  Web}, {{WWW}} '16, pages 771--782. {International World Wide Web Conferences
  Steering Committee}, 2016.
\newblock ISBN 978-1-4503-4143-1.

\bibitem[Unger et~al.(2012)Unger, B{\"u}hmann, Lehmann, Ngonga~Ngomo, Gerber,
  and Cimiano]{Unger12}
C.~Unger, L.~B{\"u}hmann, J.~Lehmann, A.-C. Ngonga~Ngomo, D.~Gerber, and
  P.~Cimiano.
\newblock Template-based question answering over {{RDF}} data.
\newblock In \emph{Proc. of the 21st International Conference on World Wide
  Web}, {{WWW}} '12, pages 639--648, {Lyon, France}, 2012. {ACM}.
\newblock ISBN 978-1-4503-1229-5.

\bibitem[Vakulenko and Savenkov(2017)]{vakulenkoS17}
S.~Vakulenko and V.~Savenkov.
\newblock {{TableQA}}: {{Question Answering}} on {{Tabular Data}}.
\newblock \emph{CoRR}, abs/1705.06504, 2017.

\bibitem[Vieira et~al.(2006)Vieira, da~Silva, Pinto, de~Moura, Cavalcanti, and
  Freire]{vieira2006FastRobust}
K.~Vieira, A.~S. da~Silva, N.~Pinto, E.~S. de~Moura, J.~M.~B. Cavalcanti, and
  J.~Freire.
\newblock A fast and robust method for web page template detection and removal.
\newblock In \emph{Proceedings of the 15th {ACM} international conference on
  {Information} and knowledge management}, {CIKM} '06, pages 258--267, New
  York, NY, USA, Nov. 2006. Association for Computing Machinery.
\newblock ISBN 978-1-59593-433-8.
\newblock \doi{10.1145/1183614.1183654}.
\newblock URL \url{https://doi.org/10.1145/1183614.1183654}.

\bibitem[Voorhees and Tice(2000)]{Voorhees00}
E.~M. Voorhees and D.~M. Tice.
\newblock Building a question answering test collection.
\newblock In \emph{Proc. of the 23rd Annual International {{ACM SIGIR}}
  Conference on Research and Development in Information Retrieval}, {{SIGIR}}
  '00, pages 200--207, {Athens, Greece}, 2000. {ACM}.
\newblock ISBN 1-58113-226-3.

\bibitem[Xiong et~al.(2016)Xiong, Zhong, and Socher]{caiming16}
C.~Xiong, V.~Zhong, and R.~Socher.
\newblock {Dynamic Coattention Networks For Question Answering}.
\newblock \emph{CoRR}, abs/1611.01604, 2016.

\bibitem[Xiong et~al.(2018)Xiong, Zhong, and Socher]{dcn18}
C.~Xiong, V.~Zhong, and R.~Socher.
\newblock {DCN+: Mixed Objective And Deep Residual Coattention for Question
  Answering}.
\newblock In \emph{Proc. of the 6th International Conference on Learning
  Representations, {ICLR} 2018}, 2018.

\bibitem[Xu et~al.(2020)Xu, Campagna, Li, and Lam]{schema2qa}
S.~Xu, G.~Campagna, J.~Li, and M.~S. Lam.
\newblock {{Schema2QA}}: {{Answering}} complex queries on the structured web
  with a neural model.
\newblock \emph{arXiv preprint arXiv:2001.05609}, 2020.

\bibitem[Yahya et~al.(2012)Yahya, Berberich, Elbassuoni, Ramanath, Tresp, and
  Weikum]{Yahya12}
M.~Yahya, K.~Berberich, S.~Elbassuoni, M.~Ramanath, V.~Tresp, and G.~Weikum.
\newblock Natural language questions for the web of data.
\newblock In \emph{Proc. of the 2012 Joint Conference on Empirical Methods in
  Natural Language Processing and Computational Natural Language Learning},
  {{EMNLP}}-{{CoNLL}} '12, pages 379--390. {ACL}, {ACL}, 2012.

\bibitem[Yi et~al.(2003)Yi, Liu, and Li]{Yi03}
L.~Yi, B.~Liu, and X.~Li.
\newblock Eliminating noisy information in web pages for data mining.
\newblock In \emph{Proc. of the Ninth {{ACM SIGKDD}} International Conference
  on Knowledge Discovery and Data Mining}, {{KDD}} '03, pages 296--305,
  {Washington, D.C.}, 2003. {ACM}.
\newblock ISBN 1-58113-737-0.

\bibitem[Zhang et~al.(2016)Zhang, Liu, He, Ji, Liu, Wu, and Zhao]{Zhang16}
Y.~Zhang, K.~Liu, S.~He, G.~Ji, Z.~Liu, H.~Wu, and J.~Zhao.
\newblock Question answering over knowledge base with neural attention
  combining global knowledge information.
\newblock \emph{CoRR}, abs/1606.00979, 2016.

\bibitem[Zou et~al.(2014)Zou, Huang, Wang, Yu, He, and Zhao]{Zou14}
L.~Zou, R.~Huang, H.~Wang, J.~X. Yu, W.~He, and D.~Zhao.
\newblock Natural language question answering over {{RDF}}: {{A}} graph data
  driven approach.
\newblock In \emph{Proc. of the 2014 ACM International Conference on Management
  of Data}, {{SIGMOD}} '14, pages 313--324. {ACM}, 2014.
\newblock ISBN 978-1-4503-2376-5.

\end{thebibliography}
}

%
%
\end{document}